\documentclass{article}
\usepackage[utf8]{inputenc}
\usepackage[margin=0.8in]{geometry}
\usepackage{amsmath,amsfonts,amsthm,amssymb}
\usepackage{bbm}

\newcommand{\eps}{\ensuremath{\varepsilon}}

\newcommand{\patrasku}{P\v{a}tra\c{s}cu}

\newcommand{\bit}{\ensuremath{\{0, 1\}}}

\newcommand{\blocks}[2]{\ensuremath{ #1_{1}, \ldots, #1_{#2} }}

\newcommand{\indicate}[1]{\ensuremath{\mathbbm{1}\{{#1}\}}}

\newcommand{\numberRep}[1]{v_{#1}}
\newcommand{\butterflyEdge}[3]{e_{#1}(#2, #3)}
\newcommand{\groupSize}{\lvert G \rvert}
\newtheorem{theorem}{Theorem}
\newtheorem{definition}{Definition}

\newtheorem{conjecture}{Conjecture}
\newtheorem{lemma}{Lemma}

\renewcommand{\log}{\lg}

\title{Stronger 3SUM-Indexing Lower Bounds}
\author{Eldon Chung\thanks{{\tt echung.math@gmail.com}. Center for Quantum Technologies, National University of Singapore.}  \and Kasper Green Larsen\thanks{{\tt larsen@cs.au.dk}. Aarhus
    University. Supported by Independent Research Fund Denmark (DFF) Sapere Aude Research Leader grant
No 9064-00068B.}}
\date{}

\begin{document}

\maketitle
\begin{abstract}
    The $3$SUM-Indexing problem was introduced as a data structure version of the $3$SUM problem, with the goal of proving strong conditional lower bounds for static data structures via reductions. Ideally, the conjectured hardness of $3$SUM-Indexing should be replaced by an unconditional lower bound. Unfortunately, we are far from proving this, with the strongest current lower bound being a logarithmic query time lower bound by Golovnev et al. from STOC'20. Moreover, their lower bound holds only for non-adaptive data structures and they explicitly asked for a lower bound for adaptive data structures. Our main contribution is precisely such a lower bound against adaptive data structures. As a secondary result, we also strengthen the non-adaptive lower bound of Golovnev et al. and prove strong lower bounds for $2$-bit-probe non-adaptive $3$SUM-Indexing data structures via a completely new approach that we find interesting in its own right.
\end{abstract}
\section{Introduction}
In the $3$SUM Problem, we are given a set $S$ of $n$ group elements from an abelian group $(G,+)$ and the goal is to determine whether there is a triple $a,b,c \in S$ such that $a+b = c$. The $3$SUM Problem was originally introduced by Gajentaan and Overmars~\cite{3SUM} as a means of establishing hardness of geometric problems. Concretely, it was conjectured that $3$SUM requires $\Omega(n^2)$ time when the underlying group is the set of reals and we use the Real-RAM computational model. By reductions, this conjecture implies similar lower bounds for a wealth of geometric problems, see e.g.~\cite{Geom3SUM1,Geom3SUM2}.

While originally being restricted mostly to geometric problems, the seminal work by \patrasku~\cite{TowardsPoly} showed that a suitable integer version of $3$SUM (e.g. $G$ is the integers modulo $n^3$), may be used to prove hardness of numerous fundamental algorithmic problems (see e.g.~\cite{Higher3SUM,Jumbled,WeightSubgraph,TowardsPoly}) in the more realistic word-RAM model. These lower bounds are based on the so-called $3$SUM Conjecture, asserting that no $n^{2-\delta}$ time $3$SUM algorithm exists for any constant $\delta>0$. To date, the fastest $3$SUM algorithm runs in time $O(n^2 (\lg \lg n)^{O(1)}/\lg^2 n)$~\cite{Fast3SUM}, which is far from refuting the conjecture. The $3$SUM Conjecture is now one of the pillars in fine-grained complexity and much effort has gone into understanding its implications for algorithm lower bounds. 

Highly related to algorithm lower bounds is lower bounds for data structures. While more progress has been made on proving unconditional lower bounds for data structures compared to algorithms, current state-of-the-art lower bounds are still only polylogarithmic~\cite{DynamicLB,ButterDynamic,Static}. This lack of progress motivates fine-grained conditional lower bounds also for data structures. The first approach in this direction, is via the Online Matrix-Vector Problem by Henzinger et al.~\cite{OMV}. Their framework yields polynomial conditional lower bounds for dynamic data structures via reductions from multiplication of a boolean matrix and a boolean vector, with addition replaced by OR and multiplication replaced by AND. However, their framework is inherently tied to \emph{dynamic} data structure problems, where a data set is to be maintained under update operations. As a means to addressing \emph{static} data structure problems, Goldstein, Kopelowitz, Lewenstein, and Porat in \cite{GKLP17} introduced the $3$SUM-Indexing Problem.

\paragraph{$3$SUM-Indexing.}
The $3$SUM-Indexing problem was first defined by Demaine and Vadhan in an unpublished manuscript \cite{DV01} and then by Goldstein, Kopelowitz, Lewenstein, and Porat in \cite{GKLP17} and is as follows:
\begin{definition}[$3$SUM-Indexing]\label{defn:3sumIndexing}
    Let $(G, +)$ be a finite abelian group. Preprocess two sets of group elements $A_1, A_2 \subseteq G$ each of size $n$ into a data structure of $S$ memory cells of $w$ bits so that given any query group element $z$, deciding whether there exists $a_1 \in A_1$ and $a_2 \in A_2$ such that $a_1 + a_2 = z$ is done by accessing at most $T$ memory cells.
\end{definition}

A number of hardness conjectures were provided together with the definition of the $3$SUM-Indexing Problem. Combined with reductions, these conjectures allow establishment of conditional lower bounds for static data structures. To be consistent with the terminology used for unconditional data structure lower bounds, which are typically proved in the cell probe model~\cite{CellProbe}, we refer to accessing a memory cell as \emph{probing} the cell. The following conjectures were made regarding the hardness of $3$SUM-Indexing:

\begin{conjecture}[\cite{GKLP17}]
\label{cnj:a}
Any data structure for $3$SUM-Indexing with space $S$ and $T=O(1)$ probes must have $S = \tilde{\Omega}(n^2)$.
\end{conjecture}

\begin{conjecture}[\cite{DV01}]
\label{cnj:b}
Any data structure for $3$SUM-Indexing with space $S$ and $T$ probes must have $ST = \tilde{\Omega}(n^2)$.
\end{conjecture}

\begin{conjecture}[\cite{GKLP17}]
\label{cnj:c}
Any data structure for $3$SUM-Indexing with space $S$ and $T=O(n^{1-\delta})$ probes must have $S = \tilde{\Omega}(n^2)$.
\end{conjecture}
Clearly the last conjecture is the strongest, and in general, we have the following implications:
\[
\textrm{Conjecture~\ref{cnj:c}} \Rightarrow \textrm{Conjecture~\ref{cnj:b}}  \Rightarrow \textrm{Conjecture~\ref{cnj:a}} 
\]
These conjectures have been successfully used to prove fine-grained hardness of several natural static data structure problems ranging from Set Disjointness, Set Intersection, Histogram Indexing to Forbidden Pattern Document Retrieval~\cite{GKLP17}. 

Very surprisingly, Golovnev et al.~\cite{GGHPV20} showed that the strongest of these conjectures, Conjecture~\ref{cnj:c}, is false. Concretely, they gave a data structure for $3$SUM-Indexing with $T = \tilde{O}(n^{3\delta})$ and $S = \tilde{O}(n^{2-\delta})$ for any constant $\delta>0$. This refutes Conjecture~\ref{cnj:c}, but not the remaining two conjectures. Their data structure is based on an elegant use of Fiat and Naor's~\cite{FuncInv} general time-space tradeoff for function inversion.

The refutation of Conjecture~\ref{cnj:c} only makes it more urgent that we replace these conjectured lower bounds by unconditional ones. However, depressingly little is still known in terms of unconditional hardness of $3$SUM-Indexing. First,~\cite{DV01} proved Conjecture~\ref{cnj:a} in the special case of $T=1$. Secondly, in the recent work by Golovnev et al.~\cite{GGHPV20}, the following was proved for \emph{non-adaptive} data structures:
\begin{theorem}[\cite{GGHPV20}]
\label{thm:previous}
    Any \emph{non-adaptive} cell probe data structure answering $3$SUM-Indexing queries for input sets of size $n$ from an abelian group $G$ of size $O(n^2)$ using $S$ words of $w$ bits must have query time $T = \Omega( \log n / \log(Sw/n) )$.
\end{theorem}
A non-adaptive data structure is one in which the cells to probe are chosen beforehand as a function \emph{only} of the query element $z$. That is, the data structure is not allowed to choose which memory cells to probe based on the contents of previously probed cells. Proving lower bounds for non-adaptive data structures is often easier than allowing adaptivity, see e.g.~\cite{AdaptOrDie,NonAdaptPred,NonAdaptSunflower}, and Golovnev et al. remark:
"\emph{It is crucial for our proof that the input is chosen at random after the subset of data structure cells, yielding a lower bound only for non-adaptive algorithms}."~\cite{GGHPV20}. Golovnev et al. explicitly raised it as an interesting open problem (Open Question 3 in~\cite{GGHPV20}) whether a similar lower bound can be proved also for adaptive data structures.

\subsection{Our Contributions}
Our main contribution is a lower bound for $3$SUM-Indexing that holds also for adaptive data structures:
\begin{theorem}
\label{thm:main}
    Any cell probe data structure answering $3$SUM-Indexing queries for input sets of size $n$ for abelian groups $([m], + \mod m)$ with $m = O(n^2)$ and $(\bit^{2\log(n)+O(1)}, \oplus)$ using $S$ words of $w = \Omega(\lg n)$ bits must have query time $T = \Omega( \log n / \log(Sw/n) )$.
\end{theorem}
Our lower bound matches the previous bound from~\cite{GGHPV20}, this time however allowing adaptivity. Moreover, it (essentially) matches the strongest known lower bounds for static data structures (the strongest lower bounds peak at $T = \Omega(\log n/\log(Sw/n))$~\cite{Static}), thus ruling out further progress without a major breakthrough (also in circuit complexity~\cite{ViolaBarrier,DvirBarrier}).

Our proof is based on a novel reduction from \patrasku's Reachability Oracles in the Butterfly graph problem~\cite{P11}. This problem, while rather abstract, has been shown to capture the hardness of a wealth of static data structure problems such as 2D Range Counting, 2D Rectangle Stabbing, 2D Skyline Counting and Range Mode Queries, see e.g.~\cite{ButterDistance,ButterSkyline,ButterMode} as well as for dynamic data structure problems, including Range Selection and Median~\cite{ButterDynamic} and recently also all dynamic problems that the Marked Ancestor Problem reduces to~\cite{FurtherUnifying,MarkedAncestor}, which includes 2d Range Emptiness, Partial Sums and Worst-Case Union-Find. Our work adds $3$SUM-Indexing and all problems it reduces to, to the list.

\paragraph{Even Smaller Universes.}
The reduction from Reachability Oracles in the Butterfly Graph problem gives lower bounds for abelian groups of size $\Omega(n^2)$, leaving open the possibility of more efficient data structures for smaller groups. Indeed, $\Omega(n^2)$ cardinality of the groups seems like a natural requirement for hardness, as there are $n^2$ pairs of elements $a_1 \in A_1$ and $a_2 \in A_2$ and thus for smaller groups, one might start to exploit structures in the sumset $A_1 + A_2$ to obtain more efficient data structures. We therefore investigate whether the lower bound in Theorem~\ref{thm:main} can be generalized to smaller groups. Quite surprisingly, we show that:
\begin{theorem}
\label{thm:smalluni}
    Any cell probe data structure answering $3$SUM-Indexing queries for input sets of size $n$ for abelian groups $([m], + \mod m)$, with $m = O(n^{1+\delta})$ and $(\bit^{(1+\delta)\log(n)+O(1)}, \oplus)$ for a constant $\delta>0$, using $S$ words of $w = \Omega(\lg n)$ bits must have query time $T = \Omega( \log n / \log(Sw/n) )$.
\end{theorem}
Thus we get logarithmic lower bounds for linear space data structures, even when the group has size only $n^{1+\delta}$.

To prove Theorem~\ref{thm:smalluni}, we revisit \patrasku's Lopsided Set Disjointness (LSD) communication game, which he also used to prove his lower bound for Reachability Oracles in the Butterfly graph problem. We give a careful reduction from LSD to $3$SUM-Indexing on small universes, thereby establishing Theorem~\ref{thm:smalluni}.

\paragraph{Non-Adaptive Data Structures.}
As another contribution, we revisit the non-adaptive setting considered by Golovnev et al.~\cite{GGHPV20}. Here we present a significantly shorter proof of their lower bound and also improve it from $T=\Omega(\lg n/\lg(Sw/n))$ to $T=\Omega(\lg |G|/\lg(Sw/n))$. Concretely, we prove the following theorem:
\begin{theorem}
\label{thm:nonadaptive}
    Any non-adaptive cell probe data structure answering $3$SUM-Indexing queries for input sets of size $n$ for an abelian group $G$ of size $\omega(n^2)$, using $S$ words of $w = \Omega(\lg n)$ bits must have query time $T = \Omega(\min\{ \log |G| / \log(Sw/n), n/w\})$.
\end{theorem}
We remark that the proof of Golovnev et al.~\cite{GGHPV20} cannot be extended to a $\lg |G|$ (technically, they require $|G|/n$ queries to survive a cell sampling, whereas we only require $n$ queries to survive). 

Our improvement has a subtle, but interesting consequence. Concretely, if the size of the group grows to sub-exponential in $n$, say $|G| = 2^{\sqrt{n}}$, then the lower bound becomes $T=\Omega(\min\{ \sqrt{n} / \log(\frac{Sw}{n}), n/w\})$. Since it is most natural to assume the cell size is large enough to store a group element, i.e. $w = \Omega(\lg |G|) = \Omega(\sqrt{n})$, the lower bound is still at least $T = \Omega(\sqrt{n}/\lg S)$. While such large groups are perhaps unrealistic, one can also interpret the result as saying that if we are non-adaptive and attempt to design a data structure that does not exploit the size of the underlying group, then we are doomed to have a slow query time.

\paragraph{Non-Adaptive $2$-Bit-Probe Data Structures.}
Finally, we consider non-adaptive data structures restricted to $T=2$ probes in the \emph{bit probe model}, meaning that each memory cell has $w=1$ bits. The lower bound from Theorem~\ref{thm:previous} by~\cite{GGHPV20} in this case is $S = \tilde{\Omega}(n^{3/2})$ (see the paper~\cite{GGHPV20} for the general formulation $S = \tilde{\Omega}(n^{1+1/T})$) and our lower bound from Theorem~\ref{thm:nonadaptive} is $S = \tilde{\Omega}(n (|G|/n)^{1/T}) = \tilde{\Omega}(\sqrt{n |G|})$. We significantly strengthen this result by proving an $S = \Omega(|G|)$ lower bound for an abelian group $(G,+)$, completely ruling out any non-trivial data structure with $2$ non-adaptive bit probes (with $|G|$ space, we can trivially store a bit vector representing the sumset $A_1 + A_2$ and have $T=1$ while being non-adaptive):
\begin{theorem}
\label{thm:bitprobe}
    Any non-adaptive data structure for $3$SUM-Indexing such that $T = 2$ and $w = 1$ requires $S = \Omega(|G|)$ for an abelian group $(G,+)$.
\end{theorem}
Our proof takes an interesting new approach to data structure lower bounds and we find that the proof itself is a valuable contribution to data structure lower bounds. The basic idea is to view the memory cells of the data structure as a graph with one node per cell. The queries then become edges corresponding to the $T=2$ memory cells probed. If the number of memory cells is $o(|G|)$, then the graph has a super-linear number of edges. This implies that its girth is at most logarithmic and hence we can find a short cycle in the graph. A cycle is a set of $m$ queries being answered by $m$ memory cells. The standard cell sampling lower bounds (often used in data structure lower bounds) cannot derive a contradiction from this, as the $m$ memory bits intuitively are sufficient to encode the $m$ query answers. However, our novel contribution is to examine the different types of possible query algorithms (i.e. which function of the two bits probed does it compute) and argue that in all cases, such a short cycle is impossible. Directly examining the types of query algorithms has not been done before in data structure lower bounds and we find this a valuable contribution that we hope may prove useful in future work.

\section{Reduction from Reachability Oracles in the Butterfly Graph}
In this section, we give a reduction from the problem of \emph{Reachability Oracles in the Butterfly Graph} to $3$SUM-Indexing with the cyclic group and the XOR group, proving Theorem~\ref{thm:main}. In both cases, the size of the group is at most quadratic with respect to the input set sizes.

\begin{definition}[Butterfly Graphs]
    A Butterfly graph of degree $B$ and depth $d$ is a directed graph with $d + 1$ layers, each comprising of $B^d$ nodes. For each layer, the $i^{th}$ node can be associated with a $d$-digit number in base $B$ which we will refer to as its \emph{label} $\numberRep{i}$ where $\numberRep{i}[0]$ denotes the least significant digit. Then there is an edge from node $i$ on the $k^{th}$ layer to node $j$ on the $(k+1)^{th}$ layer if and only if $\numberRep{i}[h] = \numberRep{j}[h]$ for all $h \neq k$. That is to say, that there is an edge if and only if $i$ and $j$ may differ only on the $k^{th}$ digit of their labels. We will denote such an edge by $\butterflyEdge{k}{i}{j}$.
    
    Nodes in the layer $0$ of the graph are called \emph{source nodes}, whereas nodes in layer $d$ of the graph are called \emph{sink nodes}.
\end{definition}

\begin{definition}[Reachability Oracles in the Butterfly Graph]
    The problem of Reachability Oracles in the Butterfly Graph is that one has to pre-process into a data structure a subset of the edges $E$ of the butterfly graph of degree $B$ and depth $d$. Queries come in the form of $(s, t)$ and the goal is decide if there exists a path from source node $s$ to sink node $t$ using the subset of edges $E$.
\end{definition}

\patrasku\ proved the following lower bound for the problem in the cell probe model:
\begin{lemma}[Section 5 of \cite{P11}]\label{lemma:butterflyLowerBound}
    Any cell probe data structure answering reachability queries in subgraphs of the butterfly graph with degree $B$ and depth $d$, using $S$ words of $w$ bits must have query time $t = \Omega(d)$, assuming that $B = \Omega(w^2)$ and $\log(B) = \Omega(\log(Sd/N))$ where $N = d B^d$.
\end{lemma}

A few remarks about reachability in the Butterfly graph are in order. Firstly, note that for any source-sink pair $(s, t)$, there exists a unique path from source $s$ to sink $t$ in the Butterfly graph. Namely, the path uses exactly edges of the form $\butterflyEdge{k}{i}{j}$ such that for $k \in [d]$, $\butterflyEdge{k}{i}{j}$ is the edge from node $i$ on layer $k$ to node $j$ on layer $k + 1$ such that:
\begin{enumerate}
    \item $\numberRep{s}[h] = \numberRep{i}[h]$ for all $h \geq k$. That is to say that the $d - k$ most significant digits of the labels of nodes $s$ and $i$ are the same. 
    \item $\numberRep{t}[h] = \numberRep{j}[h]$ for all $h \leq k$. That is to say that the $k + 1$ least significant digits of the labels of nodes $t$ and $j$ are the same. 
\end{enumerate}
Conversely, we can also say that the edge $\butterflyEdge{k}{i}{j}$ connects all pairs of nodes $s, t$ such that the label for $s$ shares the most significant $d - k$ digits with $i$ and the label for $t$ shares the least significant $k + 1$ digits with $t$. 

Intuitively, this is because the traversing from node $i$ in the $k^{th}$ layer to node $j$ in the $(k+1)^{th}$ layer can be seen as ``setting'' the $k^{th}$ digit of the label for node $i$ into the $k^{th}$ digit of the label for node $j$ while leaving the rest of the digits unaltered.

The general idea of the reduction to $3$SUM-Indexing is to test whether all the required edges are present when querying for $s$ and $t$. This should be done by asking one $3$SUM-Indexing query. We will design it such that a sum $z = a_1 + a_2$ exists for our query $z$ if and only if there is at least one edge missing on the path from $s$ to $t$. 

\paragraph{Constructing $A_1$.}
Our basic idea is to take every edge $\butterflyEdge{k}{i}{j}$ in the Butterfly graph and encode it into a group element $g$ in $A_1$. We construct $g$ such that its digits can be broken up into 5 blocks so that conceptually the:
\begin{enumerate}
    \item first block encodes the layer the edge is from;
    \item second block encodes the presence of edge $\butterflyEdge{k}{i}{j}$ in $E$;
    \item third block encodes the $d - k$ most significant bits of $i$ followed by $k$ zeroes;
    \item fourth block holds $d - k - 1$ zeroes followed by the $k + 1$ least significant digits of $j$;
    \item fifth block holds $2$ zeroes.
\end{enumerate}
In short, for every edge $\butterflyEdge{k}{i}{j}$, we add group element to $A_1$ whose digits are in the following form:
\begin{equation*}
    (k, \indicate{\butterflyEdge{k}{i}{j} \in E}, \numberRep{i}[d-1], \ldots, \numberRep{i}[k], \underbrace{0, \ldots, 0}_\text{d - 1}, \numberRep{j}[k], \ldots, \numberRep{j}[0], 0, 0)
\end{equation*}
where $\indicate{\butterflyEdge{k}{i}{j} \in E}$ is $1$ if $e_k(i, j) \in E$ and $0$ otherwise. Note that the Butterfly graph has $dB^d$ nodes with degree $B$, hence a total of $n=dB^{d+1}$ edges. Since $A_1$ has one element for each such edge, we have $|A_1|=n$.

\paragraph{Constructing $A_2$.}
Next, we construct the set $A_2$ of group elements such that for every $k$, it ``helps'' any group element $g$ in $A_1$, originating from an edge $\butterflyEdge{k}{i}{j}$, to sum to any value where the third block shares the $d - k$ most significant bits with $i$ and the fourth block shares the $k + 1$ least significant digits of $j$. This can be done by adding into set $A_2$ every group element such that the:
\begin{enumerate}
    \item first block holds some value $-k$;
    \item second block is zero;
    \item third block is $d - k$ zeroes followed by any possible $k$ digit value;
    \item fourth block holds any possible $d - k - 1$ digit value followed by $k + 1$ zeroes;
    \item fifth block holds any possible digit value from $[0, B - 1]$.
\end{enumerate}

Thus for $k \in [0, d - 1]$, we add any number of the following form into $A_2$:
\begin{equation*}
    (-k, 0, \underbrace{0, \ldots, 0}_{d - k} , \underbrace{\star, \ldots, \star}_{d - 1}, \underbrace{0, \ldots, 0}_{k+1}, \star, \star)
\end{equation*}
where $\star$ denotes wildcard. Note that the least significant digits is not strictly necessary but is included to enforce that the size of the sets $A_1$ and $A_2$ are the same. Observe that $|A_2| = dB^{d+1} = n = |A_1|$.

\paragraph{Different Groups.}
For the reduction to $3$SUM-indexing in the cyclic group, we will consider the set of integers in $[(d B^{d+1})^2]$. To that end, the encoding works by understanding the $2(d + 2)$ digits as specifying a mixed-radix number, where the most significant digit is in base $4d$, the second most significant digit is in base $3$ and the remaining digits are in base $B$. In which case, we can take $-k$ to be $4d-k$.

On the other hand, for the XOR group, assuming that $d$ and $B$ are powers of $2$, we can then also naturally transform each digit into their binary representation with the exception of the most significant digit whose bit representation should be based on the number's complement and the second most significant digit may be in base $2$.



\paragraph{Translating a Query.}
What remains is to explain how we answer a reachability query $(s,t)$. We will first consider the reduction for the group $([(d B^{d+1})^2], + \mod (d B^{d+1})^2)$ and subsequently argue that the same reduction basically holds for the XOR group assuming that $d$ and $B$ are powers of $2$. We claim that there exists $a_1 \in A_1$ and $a_2 \in A_2$ whose sum is
    \begin{equation*}
        z_{(s,t)} = (0, 0, \numberRep{s}[d-1], \ldots, \numberRep{s}[0], \numberRep{t}[d-1], \ldots, \numberRep{t}[0], 0, 0)
    \end{equation*}
    if and only if there does not exist a path from $s$ to $t$ in the Butterfly graph. 
    
    To see this, we first argue that for a pair $a_1 + a_2$ that could potentially sum to $z_{(s,t)}$, we need not worry about carries amongst the digits of the numbers. To see this, we start by observing that $a_1 + a_2$ must have its most significant digit equal to $0$. We claim this is only possible if $a_1$'s most significant digit is $k$ and $a_2$'s is $4d-k=-k$. To see this, observe that the second most significant digit of $a_1$ is at most $1$ and the second most significant of $a_2$ is always $0$. Since the second most significant digit is in base $3$, this means that we cannot get a carry from these digits. Now that we have established this, we observe that for all remaining digits of any valid pair $a_1$ and $a_2$ (pairs where the most significant digit in the sum is $0$), there is at most one of the elements that has a non-zero digit, hence we will not see any carries.
    
    Now assume there does not exists a path from some source node $s$ to some sink node $t$. This must mean that there exists a $k \in [0, d - 1]$ and an edge $e_k(i, j)$ not in $E$ where:
    \begin{align*}
        \numberRep{i} &= ( \numberRep{s}[d-1], \ldots, \numberRep{s}[k], \numberRep{t}[k - 1], \ldots, \numberRep{t}[0] ) \\
        \numberRep{j} &= ( \numberRep{s}[d-1], \ldots, \numberRep{s}[k+1], \numberRep{t}[k], \ldots, \numberRep{t}[0] )
    \end{align*}
    By construction, this implies that the following group element exists in the set $A_1$:
    \begin{equation*}
        (k, 0, \numberRep{s}[d-1], \ldots, \numberRep{s}[k], \underbrace{0, \ldots, 0}_\text{d - 1}, \numberRep{t}[k], \ldots, \numberRep{t}[0], 0, 0)
    \end{equation*}
    Furthermore, the following group element always exists in $A_2$:
    \begin{equation*}
        (-k, 0, 0, \ldots, 0, \numberRep{t}[k - 1], \ldots, \numberRep{t}[0], \numberRep{s}[d-1], \ldots, \numberRep{s}[k+1], 0, \ldots, 0, 0)
    \end{equation*}
    This means that the value $(0, 0, \numberRep{s}[d-1], \ldots, \numberRep{s}[0], \numberRep{t}[d-1], \ldots, \numberRep{t}[0], 0)$ is obtainable as a sum $a_1 + a_2$. If on the other hand there is a path between $s$ and $t$, then all elements in $A_1$ of the form
    \begin{equation*}
        (k, \star, \numberRep{s}[d-1], \ldots, \numberRep{s}[k], \underbrace{0, \ldots, 0}_\text{d - 1}, \numberRep{t}[k], \ldots, \numberRep{t}[0], 0)
    \end{equation*}
    must have $\star = 1$ and thus it is not possible to write $z_{(s,t)}$ as $a_1 + a_2$.
\paragraph{The XOR Group.}
    For a reduction to the XOR group setting, we consider each element coordinate-wise using their binary representations with the exception that in the first coordinate the value is represented using the number's complement representation. Using the previous remark we also assert that for any pair $a_1 \in A_1$, $a_2 \in A_2$, the only common digit that is both non-zero is the most significant digit and thus the addition being done digit-wise. For that reason, the sum behaves exactly the same way over the XOR group as it does over the cyclic group that we have defined. Thus the size of the universe and input sets $A_1, A_2$ remain unchanged and the reduction holds in the XOR group as well.

\paragraph{Analysis.}
    Now by setting $B = \frac{Sw^2}{n}$, note that $B = \Omega(w^2)$ and:
    \begin{align*}
        \log(Sd/N) \leq \log\left(\frac{S B\log(n)}{n}\right) \leq \log(S B w/n) \leq \log(B^2) = O(\log B)
    \end{align*}

    Furthermore, it holds that
    \begin{align*}
        \log(Sw/n) = \frac{1}{2} \log( (Sw/n)^2) \geq \frac{1}{2} \log(Sw^2/n) \geq \frac{1}{2} \log B
    \end{align*}

    Using Lemma \ref{lemma:butterflyLowerBound}, it then follows that for any cell-probe solution for $3$SUM-Indexing for the cyclic group $([m], + \mod m)$ where $m = O(n^2)$ and XOR group $(\bit^{2\log(n) + O(1)}, \oplus)$ any static data structure that uses $S \geq n$ cells of $w \geq \log(n)$ bits has query time $T = \Omega(d) = \Omega(\lg_B n) = \Omega( \log n / \log(Sw/n))$.


\section{Reduction from Lopsided Set Disjointness}
In this section, we prove Theorem~\ref{thm:smalluni}, establishing hardness of $3$SUM-Indexing also for abelian groups of size $\Delta = n^{1+\delta}$. For the proof, we focus on the integers modulo $\Delta$, but remark that the proof readily adapts to the XOR group as well.

For the proof, we use \patrasku's Blocked Lopsided Set Disjointness (Blocked LSD) problem. In this problem, there are two players, Alice and Bob. Bob receives as input a set $X$, which is an arbitrary subset of a universe $[N] \times [B]$. Alice receives a set $Y \subset [N] \times [B]$ with the restriction that $Y$ contains exactly one element $(i,b_i)$ for every $i=0,\dots,N-1$. The goal for Alice and Bob is to determine whether $X \cap Y = \emptyset$ while minimizing communication. The following is known regarding the communication complexity of Blocked LSD:
\begin{lemma}[Theorem 4 of \cite{P11}]\label{lemma:lsdLowerBound}
    Fix $\delta > 0$. Any communication protocol for Blocked LSD requires either Alice sending at least $\delta N \log B$ bits, or Bob sending at least $NB^{1-O(\delta)}$ bits.
\end{lemma}
The basic idea in the reduction, is to have Bob interpret his set $X$ as two input sets $A_1,A_2$ of $n=NB$ group elements to $3$SUM-Indexing (we may have $|A_1|$ and $|A_2|$ smaller than $NB$, but we can always pad with dummy elements, so we assume $n=NB$). Given a data structure $D$ for $3$SUM-Indexing, Bob then builds $D$ on this input. Alice on the other hand interprets her set $Y$ (which has cardinality $N$) as a set of $N/\ell$ queries to $3$SUM-Indexing, where $\ell$ is a parameter to be determined. The key property of the reduction, is that the answers to all $N/\ell$ queries of Alice on $D$, determines whether $X \cap Y = \emptyset$.

\paragraph{Communication Protocol.}
Assume for now that we can give such a reduction. Alice and Bob then obtains a communication protocol for Blocked LSD as follows: Let $T$ be the query time of $D$. For $i=1,\dots,T$, Alice simulates the $i$'th step of the query algorithm for each of her $N/\ell$ queries, \emph{in parallel}. This is done by asking Bob for the set of at most $N/\ell$ cells that they probe in the $i$'th step. This costs $O(\lg \binom{S}{N/\ell}) = O((N/\ell) \lg(S\ell/N))$ bits of communication by specifying the required cells as a subset of the $S$ memory cells of $D$. Bob replies with the contents of the cells, costing $((N/\ell)w)$ bits. This is done for $T$ rounds, resulting in a communication protocol where Alice sends $O((N/\ell) T \lg(S\ell/N))$ bits and Bob sends $O((N/\ell)Tw)$ bits. If we fix $B = w^4$ and $\delta$ as a small enough constant, then Lemma~\ref{lemma:lsdLowerBound} says that either Alice sends $\Omega(N \lg w)$ bits or Bob sends $\Omega(N \sqrt{B}) = \Omega(N w^2)$ bits. In our protocol, Bob's communication is $O((N/\ell)T w) = O(N Tw)$ bits. We assume $w = \Omega(\lg n)$, thus we conclude that either $NTw = \Omega(N w^2) \Rightarrow T = \Omega(\lg n)$, or Alice's communication must be $\Omega(N \lg B) = \Omega(N \lg w)$ bits. In the first case, we are done with the proof, hence we examine the latter case. Alice's communication is $O((N/\ell) T \lg(S \ell/N))$ bits, which implies $T = \Omega(\ell \lg w/\lg(S \ell /N))$. Thus to derive our lower bound, we have to argue that it suffices for Alice to answer $N/\ell$ queries for a large enough $\ell$.

\paragraph{Asking Few Queries.}
We will show that it suffices for Alice to ask $N/\ell$ queries with $\ell = \eps \lg n/\lg w$. Here $\eps > 0$ is a small constant depending on $\delta$ in the group size $\Delta = n^{1+\delta}$. Thus we get a lower bound of $T = \Omega(\lg n/\lg((S \lg n)/(N \lg w)))$. Since $N = n/B = n/w^4$, this simplifies to $T = \Omega(\lg n/\lg(Sw/n))$ as claimed in Theorem~\ref{thm:smalluni}.

Thus what remains is to show how Alice and Bob computes the input and queries. For this, they conceptually partition the universe $[N] \times [B]$ into groups $\{i \ell,\dots,(i+1)\ell-1\} \times [B]$ for $i=0,\dots,N/\ell$. Alice will ask precisely one query for each such group. Denote by $Y_i$ the subset of $Y$ that falls in the $i$'th group and denote by $X_i$ the subset of $X$ that falls in the $i$'th group. Clearly $X \cap Y = \emptyset$ if and only if $X_i \cap Y_i = \emptyset$ for all $i$. Thus Alice will use her $i$'th query to determine whether $X_i \cap Y_i = \emptyset$. 

\paragraph{Constructing $A_1$ and $A_2$.}
To support this, Bob first constructs the set $A_1$ based on his elements $X$. He examines each group $X_i$, and for every $(j,b) \in X_i$, he adds the integer $i(2B+1)^{\ell+1} + (b+1)(2B+1)^{j-i\ell}$ to $A_1$. Next, he constructs the set $A_2$. For this, he considers all vectors $Z=(b_0,\dots,b_{\ell - 1})$ for which the numbers are all between $0$ and $B$ and precisely one of them is $0$. He adds the integer $\sum_{j \in [\ell]} b_j (2B+1)^j$ to $A_2$. This completes Bob's construction of the input sets $A_1$ and $A_2$. We have $|A_1| \leq NB = n$ and $|A_2| \leq (2B+1)^\ell$.

\paragraph{Asking the Queries.}
We next describe how Alice translates her set $Y$ into queries. For each $Y_i$, she needs to construct one query $z_i$ whose answer determines whether $X_i \cap Y_i = \emptyset$. Recall that $Y_i$ is of the form $\{(i\ell,b_0),(i\ell+1,b_1),\dots,((i+1)\ell-1,b_{\ell-1})\}$. She starts by subtracting off $i \ell$ from the first index in each pair, obtaining the set $\{(0,b_0),(1,b_1),\dots,(\ell-1,b_{\ell-1})\}$. Alice now asks the query $z_i = i(2B+1)^{\ell+1} + \sum_{j=0}^{\ell-1} (b_j+1) (2B+1)^j$. 

\paragraph{Correctness.}
We claim that $z_i$ is part of a $3$SUM if and only if $X_i \cap Y_i \neq \emptyset$. To see this, observe first that to write $z_i$ as $a_1 + a_2$, it must be the case that $a_1$ was constructed from $X_i$ as otherwise we cannot obtain the $i(2B+1)^{\ell+1}$ parts of $z_i$. Next, observe that if we write the integers in base $2B+1$, then $A_2$ contains precisely every integer of the form where there is a single digit $j \in [\ell]$ that is zero and all remaining are non-zero. Also, the numbers obtained from $(j,b) \in X_i$ are of the form $i(2B+1)^{\ell  + 1} + (b+1)(2B+1)^{j-i\ell}$ and thus have exactly one non-zero digit among the first $\ell$. Since $z_i$ has exclusive non-zero digits in the first $\ell$, it follows that $z_i=i(2B+1)^{\ell+1} + \sum_{j=0}^{\ell-1} (b_j+1) (2B+1)^j$ can be written as $a_1 + a_2$ if and only if $a_1$ was obtained from a $(j,b) \in X_i$ for which $b$ is equal to $b_{j-i\ell}$. This is the case if and only if $X_i$ and $Y_i$ intersect in $(j,b)$.

\paragraph{Analysis.}
We now determine $\ell$. Recall that $B = w^4$ and observe that all possible integers are bounded by $N (2B+1)^{\ell + 2}  \leq n (2B+1)^{\ell + 2}$. If we insist on a group of size $\Delta = n^{1+\delta}$, this means we can set $\ell = \delta \lg n/\lg(2B+1) - 2 \geq \eps \lg n/\lg w$ for a sufficiently small constant $\eps>0$. This also implies that $|A_2| \leq n^\delta \leq n$ and thus completes the proof of Theorem~\ref{thm:smalluni}.
\section{Lower Bound for Non-Adaptive Data Structures}
In this section, we prove an $\Omega(\min\{\lg|G|/\lg(Sw/n),n/w\})$ lower bound for non-adaptive $3$SUM-Indexing data structures when $|G| = \omega(n^2)$. Similarly to the previous approach by Golovnev et al.~\cite{GGHPV20}, we use a cell sampling approach. 

Consider a data structure using $S$ memory cells of $w$ bits and answering queries non-adaptively in $T$ probes. Consider all subsets of $\Delta = n/(2w)$ memory cells. There are $\binom{S}{\Delta}$ such subsets. We say that a query $z$ is answered by a set of cells $C$, if all the (non-adaptively chosen) cells it probes are contained in $C$. Any query $z$ is answered by at least $\binom{S-T}{\Delta-T}$ sets of $\Delta$ cells, namely all those containing the $T$ cells probed on $z$. It follows by averaging over the $|G|$ queries that there is a set of cells $C^*$ answering at least
\[
|G|\binom{S-T}{\Delta-T}/\binom{S}{\Delta} = |G| \frac{\Delta(\Delta-1)\cdots(\Delta-T+1)}{S(S-1)\cdots(S-T+1)} \geq |G| \left(\frac{\Delta - T + 1}{S}\right)^T
\]
queries.

If $T \geq \Delta/2$, we are already done as we have proven $T = \Omega(n/w)$. Otherwise, $T \leq \Delta/2$ and thus the above is at least $|G|(\Delta/(2S))^T = |G|(n/(4Sw))^T$. If we assume for contradiction that $T = o(\lg|G|/\lg(Sw/n))$, this is at least $|G|^{1-o(1)} > n$. Let $Q$ be the group elements corresponding to an arbitrary subset of $n$ of those queries. We argue that we can construct a distribution over inputs $A_1,A_2$ such that the queries $Q$ cannot be answered from few cells, contradicting that we have answered them from $C^*$. More precisely, we show:

\begin{lemma}\label{lemma:independence}
    Let $(G, +)$ be an abelian group with $\omega(n^2)$ elements. Given any subset $Q \subseteq G$ of at most $n$ elements, there exists an input distribution $D$ of $A_1, A_2$ such that, all the events of the form $q \in (A_1 + A_2)$ (defined as $\{a_1 + a_2 : a_1 \in A_1, a_2 \in A_2\}$) for all $q$ in $Q$ is fully independent. That is, for any subset $S = \{ s_1, s_2 \ldots, s_r\}$ of $Q$ of $r$ elements, and any sequence of $r$ events $E_1, E_2, \ldots, E_r$ either of the form $s_i \in (A_1 + A_2)$ or the form $s_i \notin (A_1 + A_2)$, it holds that $\Pr[\bigwedge_{i = 1}^r E_i] = \prod_{i = 1}^r \Pr[E_i]$ . Furthermore, for any $q \in Q$, it is the case that $\Pr_{(A_1,A_2) \sim D}[q \in (A_1 + A_2)] = \frac{1}{2}$.
\end{lemma}
The proof is deferred to the end of the section.

We now use Lemma~\ref{lemma:independence} to derive a contradiction to the assumption that $T=o(\lg|G|/\lg(Sw/n))$. Concretely, we invoke the lemma with the $Q$ defined above. This implies that the answers to the queries in $Q$ has entropy $n$ bits. However, they are being answered from a fixed set of $n/2w$ cells. These cells together have $n/2$ bits. Since their addresses are fixed, their contents must uniquely determine the $n$ query answers, yielding the contradiction and hence $T = \Omega(\min\{\lg|G|/\lg(Sw/n), n/w\})$. This completes the proof of Theorem~\ref{thm:nonadaptive}. What remains is to prove Lemma~\ref{lemma:independence}:

\paragraph{Proof of Lemma \ref{lemma:independence}.}
We prove the lemma by first showing that given $Q \subseteq G$ of $n$ elements, for any $P \subseteq Q$ there exists an input pair $A_1^P$ and $A_2^P$ such that $P \subseteq (A_1^P + A_2^P)$ and $(Q \setminus P) \cap (A_1^P + A_2^P) = \emptyset$. That is to say that for every possible subset $P$ of $Q$, there exists a pair of sets $(A_1^P, A_2^P)$ such that $(A_1^P + A_2^P)$ contains all the pair sums of $P$ and none of the pair sums outside of $P$ and in $Q$. Then $D$ is the distribution that is uniform over all possible pairs of sets $(A_1^P, A_2^P)$ with $P$ ranging over all subsets of $Q$. Another way to view $D$ is the distribution that first randomly samples $P \subseteq Q$ before deterministically outputing pairs of sets $(A_1^P, A_2^P)$.

Given any $P$, we build the sets $A_1^P$ and $A_2^P$ iteratively, where they are both initially empty. Let $p_1, p_2, \ldots$ enumerate the elements of $P$. At each iteration, let $p_i$ be the first value not in $(A_1^P + A_2^P)$. There are $\lvert G \rvert$ ordered pairs of elements $(a_1, a_2)$ such that $a_1 + a_2 = p_i$. To see this, note that letting $a_1 = p_i + (-t)$ and $a_2 = t$ for any $t \in G$ yields us a distinct pair of elements for which the sum holds. We want to show that we can add $n$ pairs of elements (thus enumerating all of the elements in $P$ and beyond) without ever having any pair sum to an element in $(Q \setminus P)$. For each element $q \in (Q \setminus P)$, and each element in $a \in A_1$, there is exactly one element $b \in G$ such that $a + b = q$ (likewise for each element $a \in A_2$). Therefore, for any given $q$, there are $\lvert A_1 \rvert$ elements $b \in G$ that if added into set $A_2$, would imply that $q \in (A_1 + A_2)$ (likewise for set $A_2$). Since $\lvert Q \setminus P \rvert \leq n$, and at every iteration $\lvert A_1 \rvert = \lvert A_2 \rvert \leq n$, we have that there are at most $2n^2$ elements that cannot be added into either set $A_1$ or set $A_2$ (otherwise sets $(Q \setminus P)$ and $(A_1 + A_2)$ are no longer disjoint).

Therefore there must still exist a pair $(a_1, a_2)$ such that $a_1 + a_2 = p_i$ and $(\{a_1\} \cup A_1^P + \{a_2\} \cup A_2^P) \cap (Q \setminus P) = \emptyset$, assuming that $\groupSize = \omega(n^2)$. In the case that every element in $P$ is enumerated before we have added $n$ pairs, we can still pad with more arbitrary pairs of elements from $G$ whilst avoiding creating any element in $(Q \setminus P)$ for the same reason as laid out above.

It remains to show that our distribution $D$ indeed witnesses full independence and that each individual event occurs with probability $\frac{1}{2}$. Let $S$ be an arbitrary subset of $Q$ of size $r \leq n$. Further, let $E_i$ be either the event that $s_i \in (A_1 + A_2)$ or $s_i \notin (A_1 + A_2)$, and let $S' \subseteq S$ contain the elements $s_i$ such that $E_i$ is the event that $s_i \in (A_1 + A_2)$ (so $S \setminus S'$ is precisely the set of elements $s_i$ for which there is the event $s_i \notin (A_1 + A_2)$). In the support of $D$, there are exactly $2^n$ pairs of sets $(A_1^P, A_2^P)$, each realising a distinct subset $P \subseteq Q$ of elements such that $P \subseteq (A_1 + A_2)$ and $(Q \setminus P) \cap (A_1 + A_2) = \emptyset$. Thus, given any set $S' \subseteq S \subseteq Q$, there are $2^{n - r}$ pairs of sets $(A_1^P, A_2^P)$ each with for set $P$ such that $S' \subseteq P$ and $(S \setminus S') \subseteq (Q \setminus P)$. Thus we argue that
\begin{equation*}
    \Pr\left[ \bigwedge_{i = 1}^r E_i \right] = \frac{2^{n - r}}{ 2^{n} } = 2^{-r}.
\end{equation*}

Note that for individual events, we can take the subset $S$ to contain only a single element $q$ from $Q$ and the above argument would imply that $\Pr[ q \in (A_1, A_2) ] = \frac{1}{2}$ and that $\Pr[ q \notin (A_1, A_2) ] = \frac{1}{2}$. Thus the conclusion readily follows from the fact that
\begin{equation*}
    \prod_{i = 1}^r \Pr\left[ E_i \right] = 2^{-r}.
\end{equation*}
\section{Bit Probe Lower Bound for $3$SUM-Indexing}
In this section we give the bit probe lower bound for $3$SUM-Indexing stated in Theorem~\ref{thm:bitprobe}.

The proof idea is based on an incompressibility argument. We will inspect the way the queries are structured and construct a specific input distribution that the data structure algorithm end up using too few bits for and therefore derive a contradiction. For this, we will again use Lemma~\ref{lemma:independence} from the previous section. The key difference between this proof and the proof in the previous section, lies in how we find a set of queries answered by too few cells. Moreover, in this proof, we will derive a contradiction even with $m$ queries being answered by $m$ cells, and thus intuitively the cells actually have enough information, but yet cannot answer the queries. We start by introducing some graph theory that we need: 

\begin{lemma}\label{lemma:girthBound}[Theorem 1 of \cite{AHL02}]
    Let $(V, E)$ be a graph with $n$ nodes, average degree $d > 2$ and girth $r$. Then $n \geq 2(d - 2)^{r / 2 - 2}$.   
\end{lemma}

From Lemma \ref{lemma:girthBound} we conclude that for a graph with $o(\groupSize)$ nodes and $\groupSize$ edges, it is the case that the graph has a girth of $O(\log (n))$. To see this, note that the average degree $d$ of such a graph is $\omega(1)$ and thus it follows that for some constant $c > 1$:
\begin{eqnarray*}
    o(\groupSize) \geq 2(d - 2)^{r / 2 - 2} &\Rightarrow& \\
    o(\groupSize) \gg 2(c)^{r / 2} &\Rightarrow &\\
    r \in O(\log_c(n))\\
\end{eqnarray*}

Given any non-adaptive pre-processing algorithm with $T = 2$, $S = o(\groupSize)$, and $w = 1$, define $V$ to be the set of $S$ nodes each representing a memory cell and let $E$ be the set of edges such that an edge $e_g = (u, v)$ is in the edge set if and only there exists some group element $g \in G$ such that the querying algorithm on input $g$ accesses both memory cell $u$ and $v$. Furthermore, associate with each edge $e_g$ a function $f_g : \bit^2 \to \bit$ that defines the output behaviour of the querying algorithm upon reading the bits at node $u$ and $v$. We broadly categorise the possible functions $f_g$ into $4$ types:
\begin{enumerate}
    \item \textbf{Copy type functions}. The type of functions $f_g$ that depend only one of its two inputs. There are $4$ of such functions.
    \item \textbf{Constant type functions}. The type of functions $f_g$ that are completely independent of its two inputs. There are $2$ such functions. 
    \item \textbf{AND type functions}. The type of functions $f_g$ whose truth table is such that exactly $3$ of the $4$ possible inputs leads to the same output where the last input differs. There are $8$ such functions.
    \item \textbf{XOR type functions}. The type of functions $f_g$ that are either the XOR of its $2$ inputs or the negation of the XOR of its $2$ inputs. There are $2$ such functions. 
\end{enumerate}

Note that none of the edges can be the constant type, since this means that the querying algorithm's answer is independent of the input set. Also, by an averaging argument there is at least one type of function that $\Omega(\groupSize)$ edges are associated with. Furthermore, Lemma \ref{lemma:independence} asserts that there can be at most $2$ edges that are parallel to each other, otherwise we can construct an input distribution $D$ such that the data structure manages to use $2$ bits to encode the outcome of a random variable that has Shannon entropy at least $3$, which is a contradiction. Thus there are $\Omega(\groupSize)$ many edges that are not parallel to each other and are all of the same type. We analyse the different types separately. We start with the simplest COPY type:

\textbf{(COPY type)} Assuming that there are $\Omega(\groupSize)$ edges that are associated with the copy type function, there must exist at least one node $u$ such that $\omega(1)$ edges $e_g$ are such that the associated function $f_g$ depend only on the bit at this node. Letting $Q$ contain two such group elements, this yields a contradiction using Lemma \ref{lemma:independence} to construct a distribution over $A_1$ and $A_2$ such that the entropy of the two query answers in $Q$ is $2$ bits.

For the remaining types, we look for a short cycle. Using Lemma \ref{lemma:girthBound}, we get that there is a cycle of $O(\log n)$ length using only edges associated with functions of the same type. Denote by $Y$ the set of group elements $g$ such that $e_g$ is in the cycle and $\blocks{y}{t}$ enumerates the elements of $Y$ based on a traversal of the cycle. That is, the edge corresponding to $y_i$ shares endpoints with edges corresponding to $y_{i-1}$ and $y_{i+1}$, where $y_{t+1} = y_1$ and $y_0 = y_t$. We use Lemma~\ref{lemma:independence} with $Q=Y$ to get a distribution $D$ over $(A_1,A_2)$ such that the answers to queries in $Y$ are independent and they are all uniform random. We now handle the two remaining types separately.

\textbf{(AND type)}
Let $b$ be the output of $f_{y_1}$ that is only obtainable by exactly $1$ of the $4$ possible inputs. Consider the distribution $D$ conditioned on the event that $\indicate{y_1 \in (A_1 + A_2)} = b$. Since only $1$ of the $4$ inputs to $f_{y_1}$ is consistent with this output, this fixes the two input bits to $f_{y_1}$. Therefore, there are $t - 2$ bits left to encode $t - 1$ independent and fully random outputs (namely, whether $y_2, \ldots, y_t$ are in $A_1 + A_2$), which yields us the desired contradiction. 

\textbf{(XOR type)}
Let $(A_1, A_2) \sim D$ be drawn from the input distribution constructed using Lemma \ref{lemma:independence} with $Q = Y$. Let the endpoints of $y_i$ be $u_i, v_i$, such that $v_t = u_1$. Note for all $i$, it is necessarily the case that $\indicate{y_i \in (A_1 + A_2)} = u_i \oplus v_i$ or $\indicate{y_i \in (A_1 + A_2)} = 1 \oplus u_i \oplus v_i$. Then $\bigoplus_2^{t} \indicate{y_i \in (A_1 + A_2)}$ is either $u_1 \oplus v_1$ or $1 \oplus u_1 \oplus v_1$ which means that $\indicate{y_1 \in (A_1 + A_2)}$ is either $\bigoplus_2^{t} \indicate{y_i \in (A_1 + A_2)}$ or its negation. Then Lemma \ref{lemma:independence} yields the desired contradiction.

\section{Acknowledgments}
The author Eldon Chung would like to thank Thomas Tan for the helpful discussions with regards to the proof for Theorem \ref{thm:bitprobe}. Also Siyao Guo for introducing the problem to him as well as initial discussions.

\bibliographystyle{abbrv}
\bibliography{main}
\end{document}